\documentclass[journal]{IEEEtran}
\ifCLASSINFOpdf
  \usepackage[pdftex]{graphicx}
\else
  \usepackage[dvips]{graphicx}
\fi

\usepackage{cite}
\usepackage{amsmath}
\usepackage{subfig}

\begin{document}
%
\title{Wideband Glide-Symmetric Slow-Wave Structure for Millimeter-Wave Sheet Beam TWTs}
%
%
%

\author{Robert~Marosi,~\IEEEmembership{Member,~IEEE,}
        Muhammed~Zuboraj,~\IEEEmembership{Senior Member,~IEEE,}
        and~Filippo~Capolino,~\IEEEmembership{Fellow,~IEEE}
\thanks{R. Marosi and F. capolino are with the Department
of Electrical Engineering and Computer Science, University of California, Irvine, Irvine,
CA, USA.

M. Zuboraj is with Los Alamos National Laboratory, Los Alamos, NM, USA.
Corresponding author e-mail: f.capolino@uci.edu.}}

%
%

\markboth{Marosi \MakeLowercase{\textit{et. al.}}: Wideband GS SWS for mm-Wave Sheet Beam TWTs, UC~Irvine, April~2025}%
{}


\maketitle

\begin{abstract}
We introduce a slow-wave structure (SWS) for a millimeter-wave sheet-beam traveling-wave tube (TWT) with wide bandwidth. The wideband and stable operation is enabled through the topological properties associated with glide-symmetry that close the bandgap at the $3\pi$-point and also make the on-axis interaction impedance negligible for the backward wave. This space harmonic structure is designed to operate in the $V$-band over 55-68 GHz with synchronism to a 5.2 kV, 11 mA sheet electron beam that will be produced by a diamond field-emitter array.
\end{abstract}


%
\IEEEpeerreviewmaketitle

\section{Introduction}

\IEEEPARstart{M}{illimeter-wave} traveling-wave tubes (TWTs) have recently become a device of interest for broadband and high-power amplification of electromagnetic (EM) waves, especially in applications such as high data rate wireless communication, imaging, and electronic countermeasures \cite{paoloni2021millimeter, armstrong2023frontiers}. At millimeter-wave and near-terahertz frequencies, microfabrication of the periodic slow-wave structures (SWSs) found in such TWTs becomes a necessity when tolerances are on the order of micrometers. Microfabrication of SWSs for TWTs has been done using various methods in literature, such as micromachining, electronic discharge machining (EDM) \cite{earley2006wideband,ives2004microfabrication,kavanagh1994evaluation}, deep reactive ion etching (DRIE) \cite{ives2004microfabrication,sengele2009microfabrication,zheng2021design}, or electroforming via the LIGA (Lithographie, Galvanoformung, Abformung) process \cite{ives2004microfabrication,han2004experimental,shin2003novel}.

Furthermore, as the operating wavelength of such devices shrinks with increasing frequency, so must the electron beam that is threaded through the SWS. As the electron beam cross-sectional area becomes small, a higher current density is required to maintain the same direct current (DC) beam power (DC accelerating voltage $V_0$ multiplied by DC beam current $I_0$). The radio frequency (RF) saturation power of TWTs is proportional to the DC beam power and depends on the basic DC-to-RF conversion efficiency, which is approximately proportional to the Pierce gain parameter $C=(Z_{\mathrm{P}}I_0/(4V_0))^{1/3}$ for $C\ll0.1$ \cite{nordsieck1953theory}. Thus, the efficiency of a TWT can be improved in three possible ways: (i) by increasing the interaction impedance ($Z_{\mathrm{P}}$) of the SWS, (ii) increasing the current of the electron beam, and/or (iii) reducing the DC acceleration voltage. 

One common strategy that we utilize in this work for increasing the beam current, efficiency, and saturation power of the TWT is to extend one dimension of the electron beam to increase its cross-sectional area (i.e., forming a rectangular or elliptical electron beam instead of a cylindrical one). Increasing the beam area allows for an increase in the total beam current without needing to use large current densities to achieve the desired DC beam power. This minimizes cathode loading (extending cathode lifetime, in the case of thermionic cathodes) and avoids the need to design complex high-perveance electron guns. Traveling-wave tubes designed with such electron beams are commonly referred to in the literature as sheet-beam TWTs \cite{ryskin2018planar,fang2018study,pershing2014demonstration,shin2011modeling,field2018development,zheng2020design,karetnikova2018gain}.

To produce the sheet beam in our TWT design, we utilize a novel diamond field emitter array (FEA) which has the same aspect ratio as our sheet electron beam. A cathode current density of approximately 5 $\mathrm{A/cm^2}$ has been prototyped using diamond FEAs \cite{ICOPS23Dkim}, thus to achieve significant beam current densities, area compression must be used. For comparison, a cathode current density as high as 1000 $\mathrm{A/cm^2}$ has been experimentally achieved in literature using a molybdenum Spindt-tip FEAs \cite{spindt1991field}, and up to 1600 $\mathrm{A/cm^2}$ with a cesiated molybdenum FEA \cite{bozler1994arrays}, which is much larger than the current density that can be achieved by our diamond FEA. However, present FEA technologies find limited practical use due to the need for a very high-quality vacuum to mitigate ion bombardment which can dull the emission points and the need for highly uniform emission current over the cross-section of the cathode (which is difficult to achieve with carbon nanotube FEAs). Thermal stresses and arcing also limit the practical use of such FEA technologies in TWTs. On the other hand, diamond field emitters show promise due to their high thermal conductivity, chemical stability, and better resilience against ion bombardment, which means that a high-quality vacuum environment will not be as critical for cathode lifetime.  In this work, we assume a rather small beam current density of 20 $\mathrm{A/cm^2}$, corresponding to a 4:1 beam area compression, assuming a cathode loading of 5 $\mathrm{A/cm^2}$. Due to the small beam current density, the saturation power that can be achieved with our current TWT design is also small, being approximately 0.724 W. For comparison, a 4.5 GHz 55 W helix TWT with a 91.4 mA pencil beam and a cathode loading of 11.5 $\mathrm{A/cm^2}$ (produced using a field emitter array), has been demonstrated experimentally in \cite{whaley2000application}. 

While increasing the interaction impedance of the SWS improves its efficiency, there is typically a trade-off between efficiency and bandwidth in TWTs. Since interaction impedance tends to decrease as wavenumber increases in space harmonic TWTs, high efficiency TWTs typically operate close to the lower band-edge frequency of the SWS. However, close to the band-edge of space harmonic TWTs, the group velocity and phase velocity vary rapidly with frequency, limiting the usable frequency range where the beam velocity and phase velocity are in synchronism and making the TWT narrowband. Here, we instead pursue the goal of wideband operation, in which the phase velocity of the wave guided by the SWS is velocity-synchronized to the electron beam over most of the passband of the SWS. Wideband operation in TWTs can generally be attributed to higher symmetry in the topology of the SWS, such as glide-symmetry (GS) or screw-symmetry, which closes bandgaps at the $3\pi$-point \cite{hessel1973propagation}. For our SWS geometry, because of topological properties associated to GS,  the on-axis interaction impedance of the backward wave vanishes, analogously to what was shown in \cite{saavedra2024wideband} for a different GS geometry, mitigating the risk of backward wave oscillations. Many geometries for wideband TWTs exhibit GS, such as in \cite{saavedra2024wideband,bian2024demonstration,castro2023wide,shin2023staggered,joye2014demonstration,pershing2014demonstration}, though GS properties are rarely acknowledged in the literature.

Here, we introduce a SWS topology for a wideband millimeter-wave sheet beam TWT, the staggered pillar SWS, with a relative bandwidth exceeding $20\%$. We utilize particle-in-cell (PIC) simulations to evaluate the bandwidth, efficiency, and saturation power of the TWT.

\begin{figure}[!t]
\centering
\subfloat(a){\includegraphics[width=0.9\columnwidth]{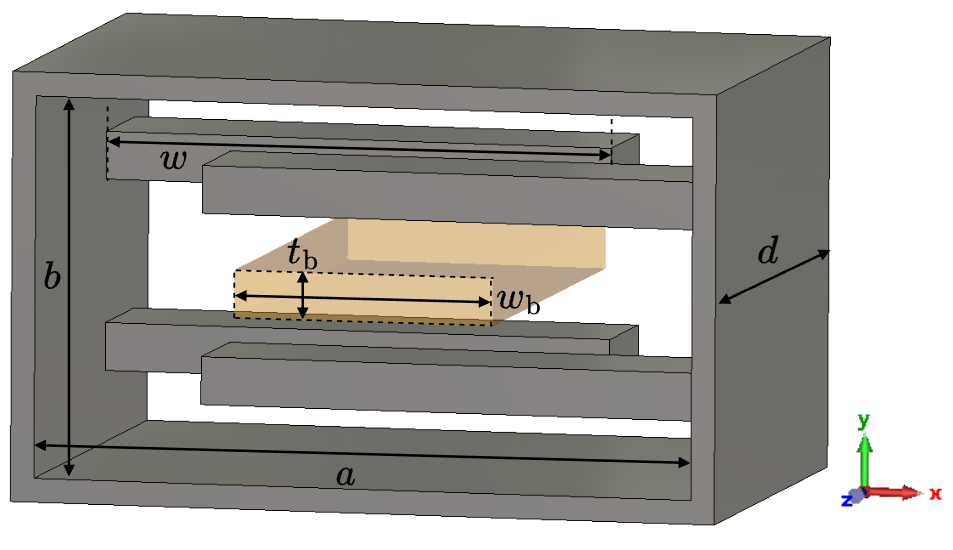}\label{Fig:unit_cell_iso}}

\subfloat(b){\includegraphics[width=0.75\columnwidth]{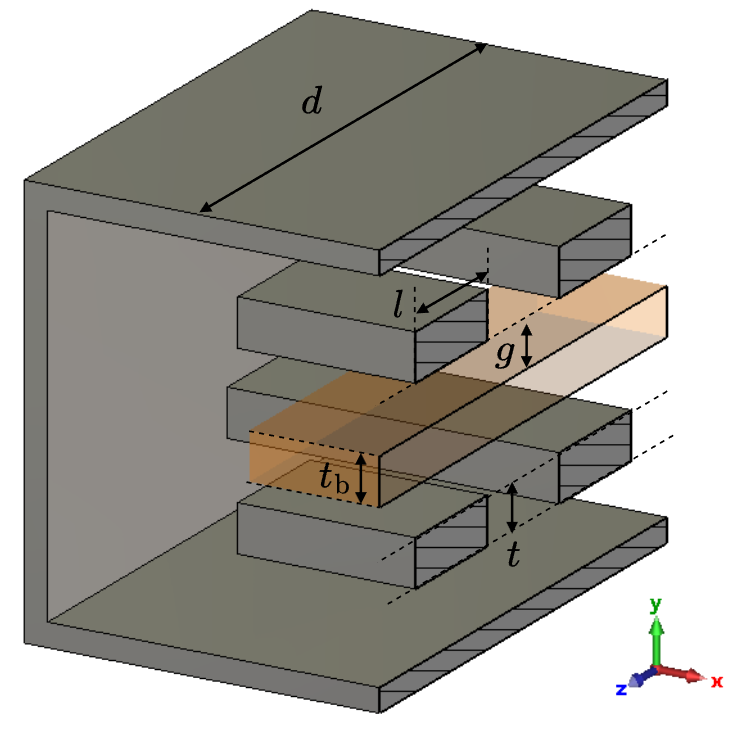}\label{Fig:unit_cell_iso_crosssec}}
\caption{Unit cell of the periodic structure with sheet electron beam in orange. (a) Isometric view; (b) longitudinal cross-section. All parts are made of metal.}
\label{fig_unit_cell}
\end{figure}

\begin{table}[]
\caption{Staggered pillar SWS unit cell dimensions}
\centering
\begin{tabular}{c||cccccc}
\hline
\textbf{Dim.} & \textbf{$a$} & \textbf{$b$} & \textbf{$d$} & \textbf{$w$} & \textbf{$l$} & \textbf{$t$} \\ 
\hline
\textbf{Val. (mm)} & 1.4 & 0.8 & 1.05 & 1.08 & 0.27 & 0.1  \\ 
\hline
\end{tabular}
\label{tab:cell_dims}
\end{table}

\section{Staggered Pillar SWS}
\begin{figure}[!t]
\centering
\includegraphics[width=1.0\columnwidth]{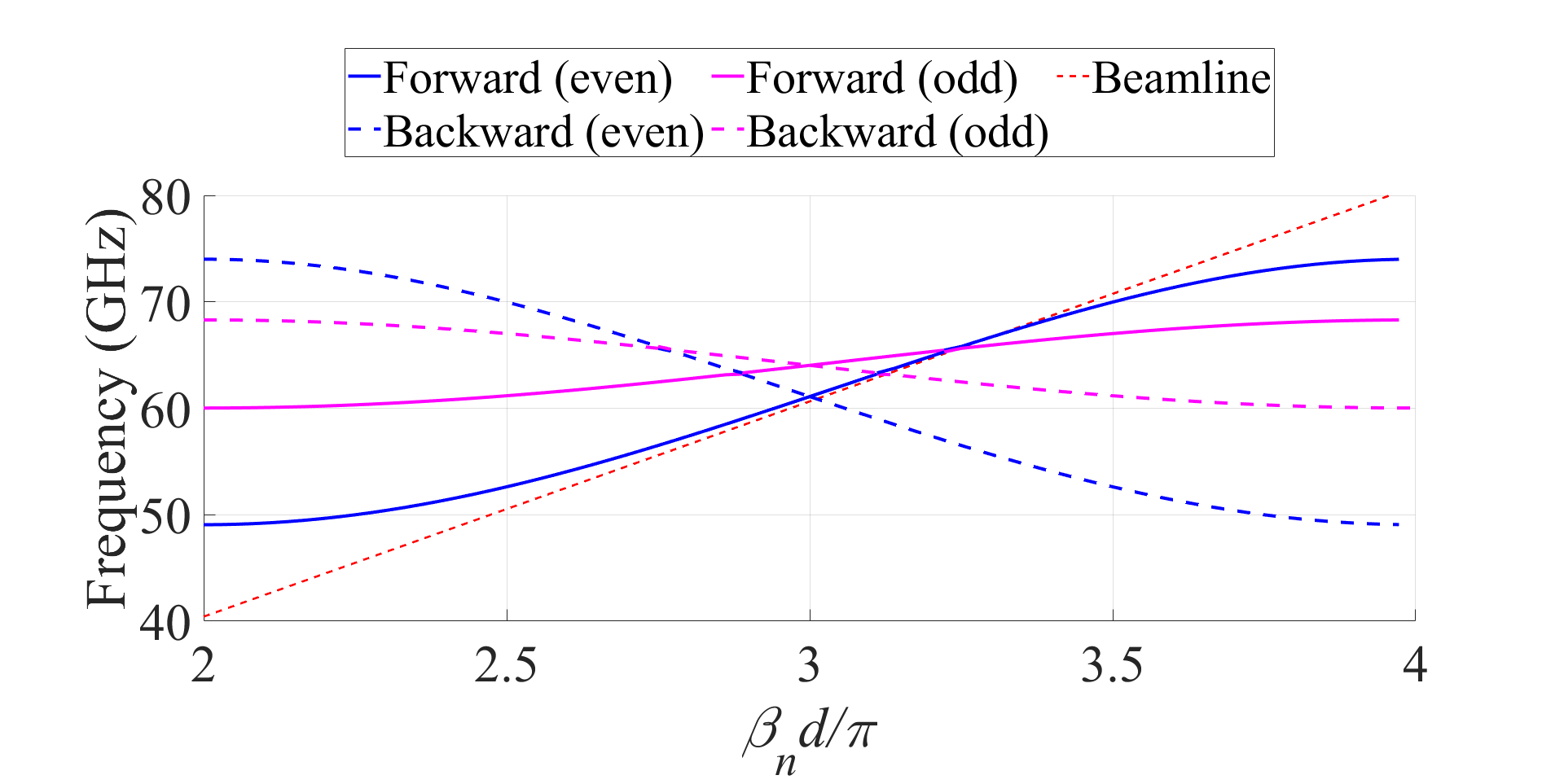}
\caption{Dispersion diagram of modes in the cold SWS. The forward  even mode (solid blue) interacts strongly with a 5.2 kV electron beam (dotted red) near $\beta_nd=3.3\pi$. The odd modes (magenta) interact weakly with beam. Forward modes are indicated by solid lines and backward modes are indicated by dashed lines. Modes cross at the $\beta_n=3 \pi/d$ point without forming a bandgap because of GS.}
\label{fig_dispersion}
\end{figure}

The staggered pillar structure consists of periodically spaced rectangular metallic pillars that extend from the narrow walls of a rectangular waveguide of height $b$ and width $a$. The SWS geometry is mirror-symmetric about the $x-z$ plane and glide-symmetric about the $y-z$ plane, as illustrated by the unit cell geometry in Fig. \ref{fig_unit_cell}. The pairs of pillars in each unit cell are spaced apart along the $z$-axis by half the cell period ($d/2$) and protrude from alternating sidewalls to make the device glide-symmetric along the $y-z$ glide plane. The structure can operate over wide bandwidths due to the property of GS that enables the modal dispersion curves of adjacent spatial harmonics to cross over each other without forming stop bands \cite{hessel1973propagation,martinez2020passband,castro2023wide} over the frequency and wavenumber interval of interest, as demonstrated in Fig. \ref{fig_dispersion}. However, tight tolerances will be required to maintain GS in practice, as breaking symmetry will lead to biperiodicity, which can result in a small bandgap at the $3\pi$-point which reduces the usable bandwidth of the device and may lead to oscillations \cite{nguyen2014design,liu2016characteristics}

The unit cell has a geometric period of $d$ along the $z$-axis. The pillars extend $w$ from the narrow walls of the rectangular waveguide and have a length of $l=d/4$ (along the $z$-axis) and thickness $t$ (along the $y$-axis). Unit cell dimensions for the staggered pillar SWS are provided in Table \ref{tab:cell_dims}. The sheet electron beam, illustrated in orange in Fig. \ref{fig_unit_cell}, has a width of $w_{\rm{b}}=0.55$ mm and a thickness of $t_{\rm{b}}=0.1$ mm. The clearance between the broad side of the pillars and the broad side of the electron beam is $g=0.1$ mm to give a good interaction impedance while also minimizing the risk of electron interception on the pillars of the SWS.

A more frequently-studied SWS topology for broadband GS sheet beam TWTs is the staggered double grating structure, which is typically synchronized to beam voltages near 20 kV \cite{shin2011modeling,field2018development,zheng2020design,karetnikova2018gain}. The staggered pillar SWS in this work has comparable bandwidth to the staggered double grating structure, and it can achieve broadband beam-wave synchronization at significantly lower beam voltages for high-efficiency operation (provided that the electron beam current can be made larger using a field emitter array).

The staggered double grating SWS is typically fabricated in GS halves that are welded together, with a seam along the $x-z$ plane, like in \cite{field2018development,zheng2020design}. Similarly, the staggered pillar SWS shown in this work can also be assembled with the weld seams along the $x-z$ plane, as explained in Appendix \ref{sec_assembly}.

Two-sided planar dielectric-supported SWSs for sheet-beam TWTs have also been studied in works such as \cite{wang2025design,nguyen2014planar,shen2012symmetric,ulisse2016w,wang2019ka}. However, with dielectric-supported structures, it is a challenge to minimize dielectric losses and charging (due to electron interception). Furthermore, planar microstrip SWSs with two central transmission lines often require complex coupler designs to only excite the desired even mode. We had initially investigated the use of planar microstrip SWSs with GS for wideband TWTs, but since microstrip SWSs such as the dielectric-supported meander line tend to be more dispersive than all-metal SWSs, wideband beam-wave synchronism due to GS could not be easily demonstrated.

\begin{figure}[!t]
\centering
\includegraphics[width=0.9\columnwidth]{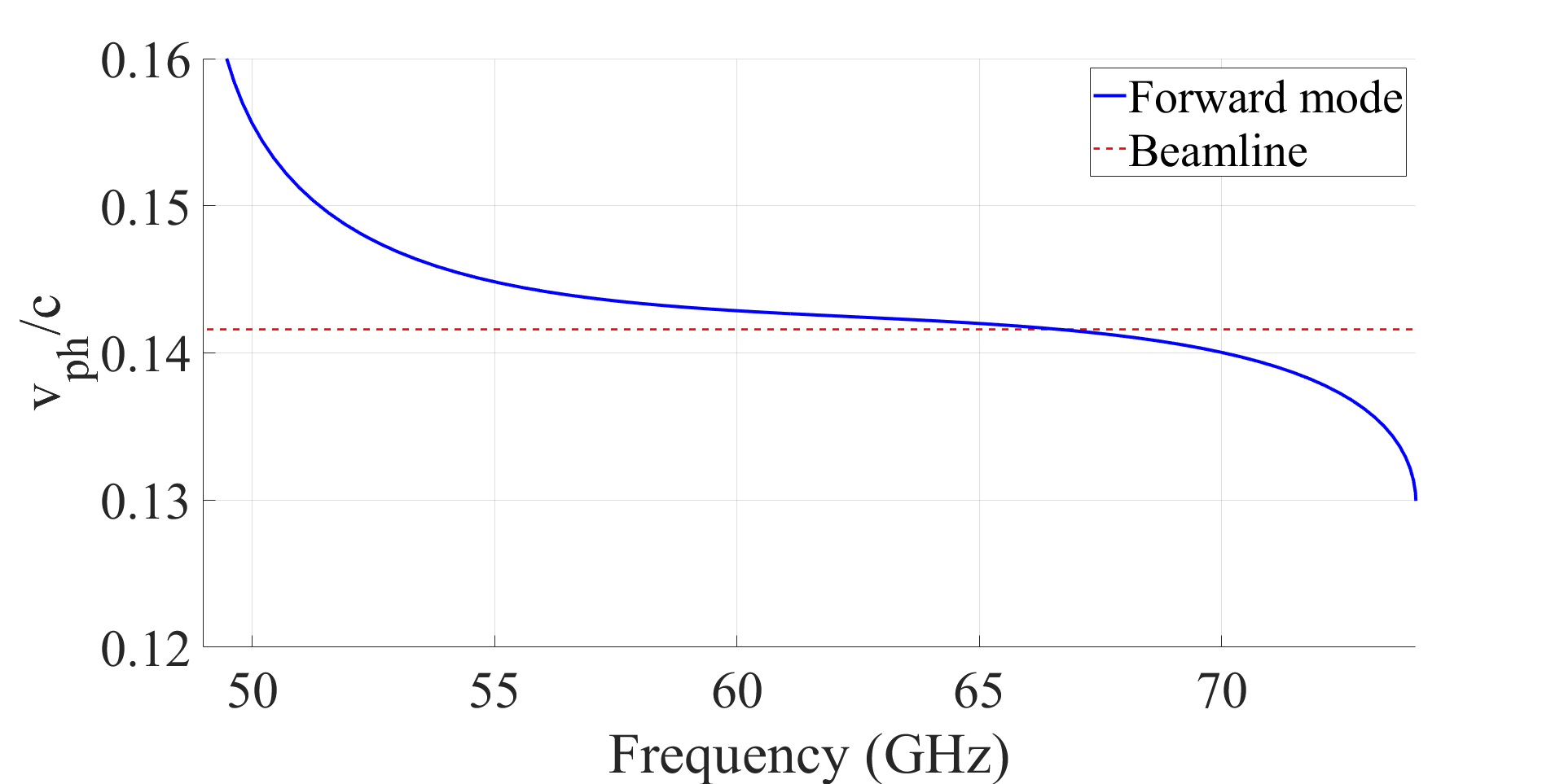}
\caption{Phase velocity of forward even mode (solid blue) over the wavenumber interval $\beta_nd=2\pi$ to $\beta_nd=4\pi$. The forward even mode synchronizes to the slow space-charge wave of a 5.2 kV electron beam (dotted red line) near $\beta_nd=3.3\pi$ and 66 GHz. This synchronization point was chosen to compensate for the smaller growth rate at high frequencies.}
\label{fig_vph}
\end{figure}

\begin{figure}[!t]
\centering
\subfloat(a){\includegraphics[width=0.9\columnwidth]{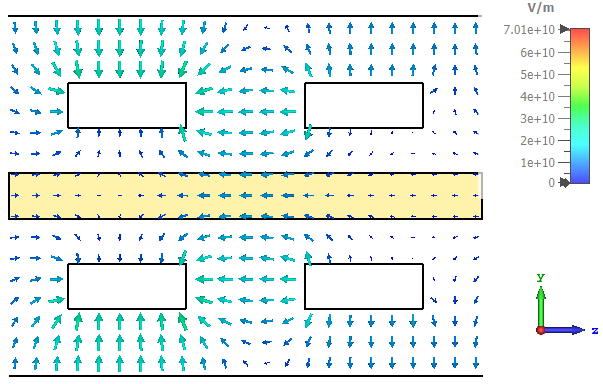}}\label{Fig:EField_even}
\subfloat(b){\includegraphics[width=0.9\columnwidth]{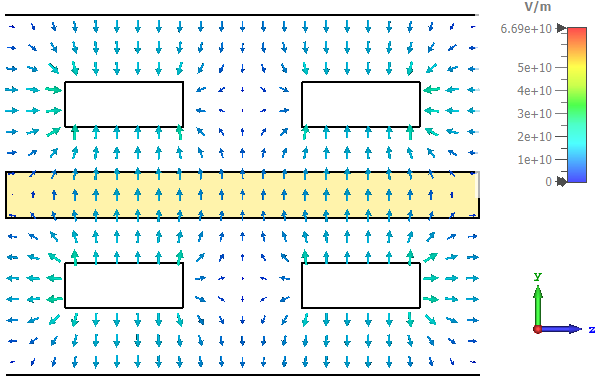}}\label{Fig:EField_odd}
\caption{(a) Electric field distribution of even eigenmode that interacts strongly with electron beam, (b) Electric field distribution of odd eigenmode that interacts weakly with electron beam slightly below $\beta_nd=3\pi$ along the longitudinal cross-section of SWS.}
\label{fig_Efield}
\end{figure}

The structure dimensions for the staggered pillar SWS were selected such that the phase velocity of the guided wave for the mode of interest (the forward-guided even mode in Fig. \ref{fig_dispersion} [solid blue], is velocity synchronized to a 5.2 kV electron beam, as shown in Fig. \ref{fig_vph}. Optimal velocity synchronization occurs above the $3\pi$-point in this design, in order to compensate for the lower growth rate (i.e., smaller interaction impedance) at high frequencies. Due to the geometry of our unit cell, both even and odd modes, with electric field distributions shown in Fig. \ref{fig_Efield}, can be supported. However, the even mode, with a strong $E_z$ component, is the mode that has the highest and most uniform interaction impedance for this design. The interaction impedance of a TWT is defined as \cite{gewartowski1965principles_ch10}

\begin{equation}
    Z_{\mathrm{P}} = \frac{|E_{z,n}|^2}{2\beta_n^2P},  \label{eqn:interaction_impedance}
\end{equation}

\noindent where $\beta_n = \beta_{\rm{c}} + 2\pi n/d$ is the phase constant of the $n$th Floquet harmonic ($n=0, \pm 1, \pm 2, ...$), $\beta_{\rm{c}}$ is the phase constant of the cold SWS evaluated within the fundamental Brillouin zone (i.e., $\beta_{\rm{c}}d/\pi\in[-1,1]$), $|E_{z,n}|$ is the longitudinal electric field of the SWS in the $n$th Floquet harmonic, and $P$ is the  time-average electromagnetic power flow for the guided mode. The TWT shown in this work is designed to operate between the first and second Floquet harmonics (i.e. $\beta_n d/\pi \in [2,4]$, as shown in Fig. \ref{fig_dispersion}), hence with $n=1$ and $n=2$.

\begin{figure}[!t]
\centering
\subfloat(a){\includegraphics[width=0.9\columnwidth]{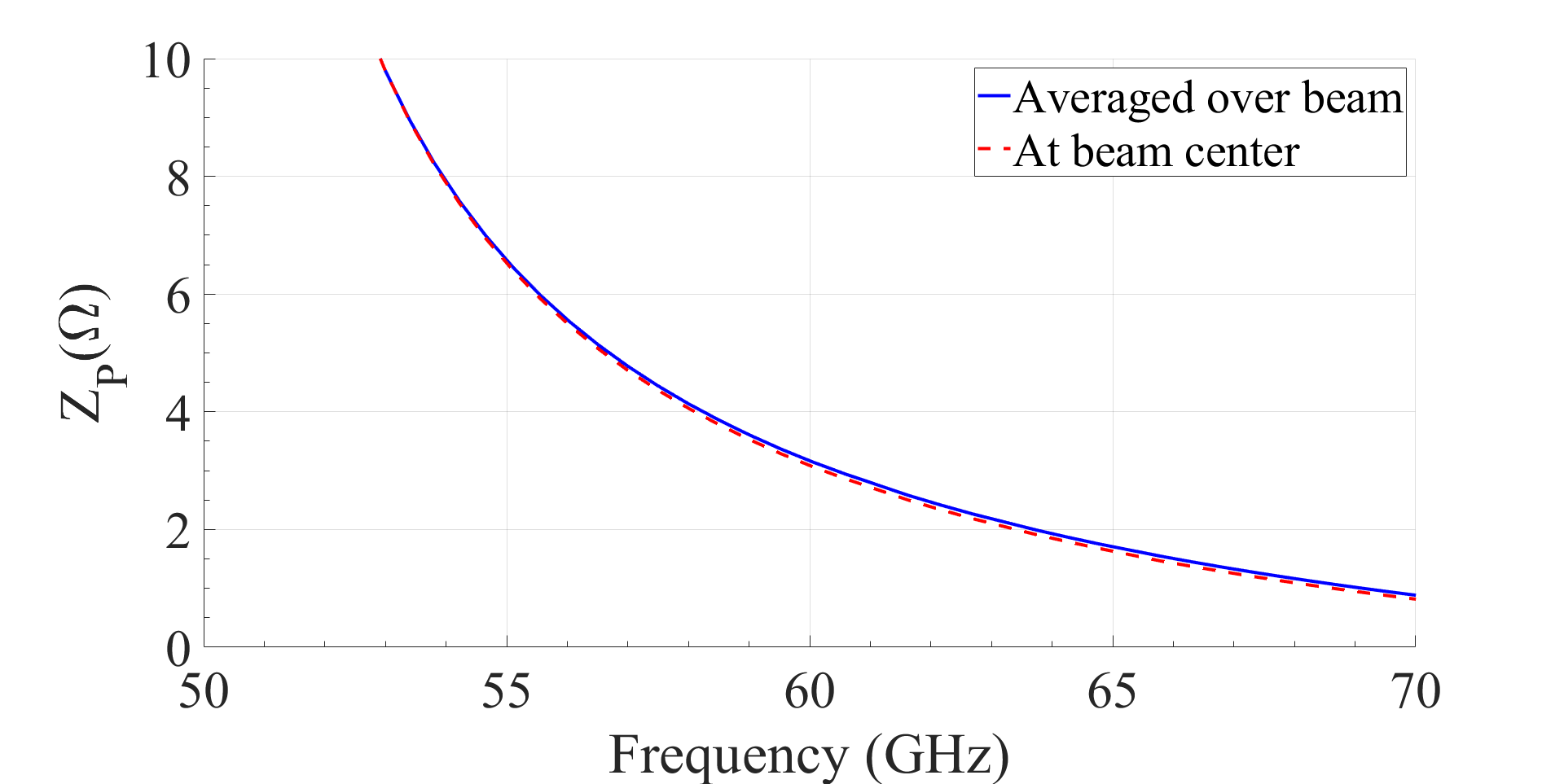}}\label{Fig:Zp_forward}
\subfloat(b){\includegraphics[width=0.9\columnwidth]{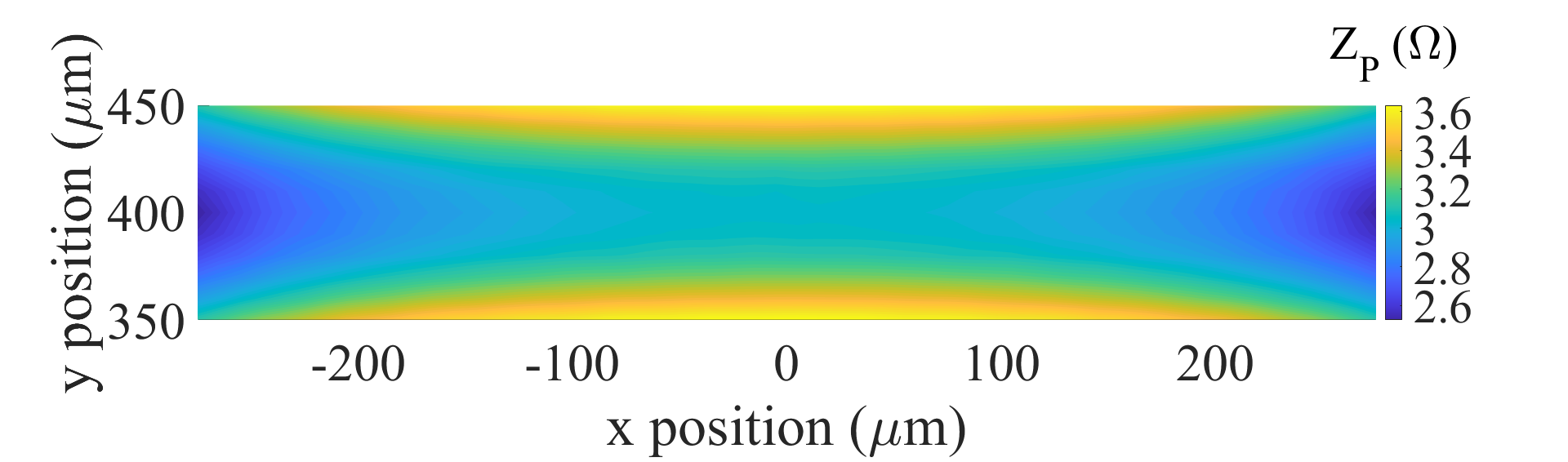}}\label{Fig:ZpContour}
\caption{(a) Interaction impedance for the forward even mode as a function of frequency averaged over the beam cross-section (solid blue) and at the center of the electron beam (dotted red). (b) Interaction impedance of forward even mode over the beam cross-section near $\beta_nd=3\pi$ and 61 GHz.}
\label{fig_Zp}
\end{figure}

The interaction impedance of the forward guided even mode at the center of the beam and averaged over the beam cross-section is shown in Fig. \ref{fig_Zp}. The interaction impedance is highly uniform over the beam cross-section, however it still varies significantly with frequency for this broadband design, with a small $Z_{\rm{P}}=1.3~\Omega$ at 68 GHz and $Z_{\rm{P}}=3.8~\Omega$ at 58 GHz. Thus, it is necessary to have optimal beam-wave synchronization at the upper frequencies in order to compensate for their small interaction impedance and to make the gain vs frequency as flat as possible. Furthermore, the coupler for our TWT, shown in Appendix \ref{sec_coupler}, is designed to only excite the even mode of the SWS. 

An important consequence of using GS is that the backward even mode supported by the SWS has an effective on-axis interaction impedance of 0 $\Omega$, as demonstrated in Appendix \ref{sec_BWO_impedance}, i.e., the longitudinal field component $E_{z,n}$ of the backward wave is absent for the spatial harmonic of interest. This important fact was previously found in \cite{castro2023wide} (however, due to the finite thickness of the electron beam, there is still a weak backward-wave interaction impedance that exists near the beam edges). Having a near-zero backward wave impedance near $\beta_nd=3\pi$ greatly mitigates the risk of backward wave oscillations, so that the TWT designer primarily needs to be concerned about regenerative oscillations (due to the combination of mismatches at the input/output ports and large gain) when checking for TWT stability.

\begin{figure}[!t]
\centering
\includegraphics[width=0.9\columnwidth]{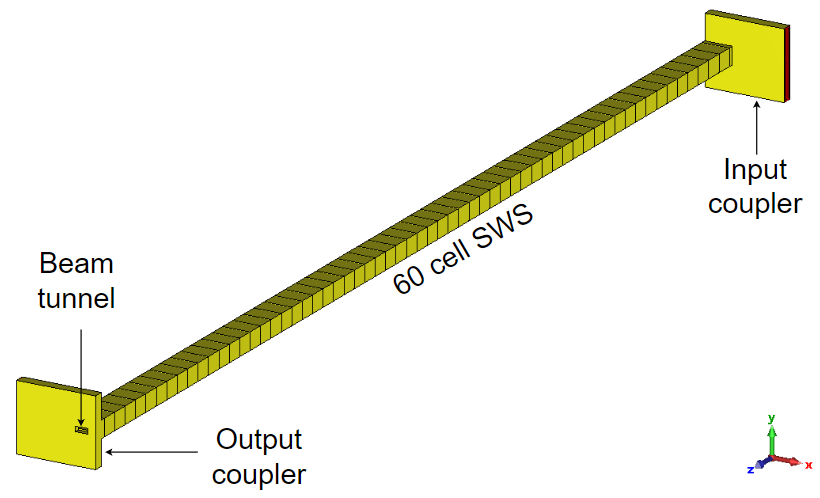}
\caption{Finite length SWS consisting of input coupler, output coupler and 60-cell interaction region.}
\label{fig_finite_length}
\end{figure}

The finite-length SWS, consisting of 60 unit cells and input/output couplers is shown in Fig. \ref{fig_finite_length}. In our simulations using the CST Studio Suite time-domain and PIC solvers, it is assumed that the SWS walls are all made of copper. The scattering parameters of the finite-length SWS and the dimensions of the input/output couplers are shown in Appendix \ref{sec_coupler}. The cold insertion loss is approximately 4.7 dB near the center frequency of the TWT and the cold -10 dB match bandwidth is approximately 16.6 GHz for the design shown in this work.

\section{Gain and PIC Results}
\begin{figure}[!t]
\centering
\includegraphics[width=0.9\columnwidth]{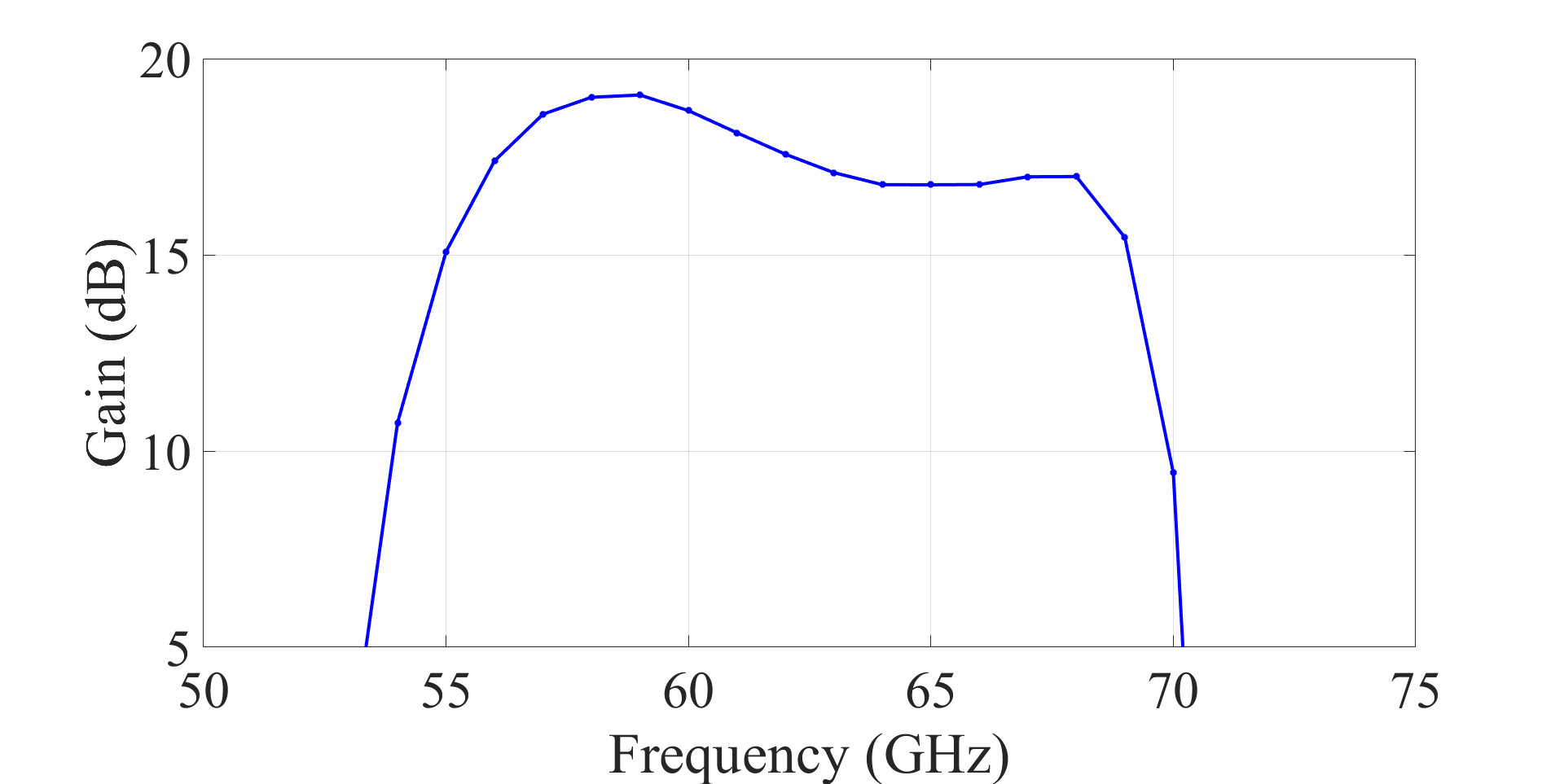}
\caption{Small-signal gain vs frequency for the TWT.}
\label{fig_gain_vs_freq}
\end{figure}

Using the PIC solver of CST Studio Suite, we evaluate the transducer power gain vs frequency and output power vs input drive curves of the TWT. Here, we define the transducer power gain of the TWT as $G=P_{\rm out}/P_{\rm av}$, where $P_{\rm av}$ is the average available power that can be delivered to the input port of the TWT (i.e. the average power of the incident wave at the input port) and $P_{\rm out}$ is the average outgoing power at the matched output port of the TWT. The PIC results were obtained using CST 2022, with 4.5 million mesh cells and 3.6 million particles. The small-signal gain of the TWT was evaluated by sweeping the frequency of a constant-wave (CW) input excitation signal with an available input power of $P_{\rm av}=1$ mW. For the purpose of demonstration, a uniform longitudinal DC magnetic field of $B_z=0.7$ T is used to confine the sheet beam. However, in practice, a periodic magnetic confinement system \cite{shi2015theoretical,booske1993stability,booske1994periodic,panda2013staggered} will likely be necessary to (i) minimize the weight and DC power consumption of the TWT, and (ii) minimize the risk of the electron beam intercepting the SWS due to $E\times B$ shear \cite{nguyen2009intense}. As shown in Fig. \ref{fig_gain_vs_freq}, the peak gain of the TWT is 19 dB at 58 GHz, with a 3 dB gain bandwidth of 13.1 GHz (22\% relative bandwidth). The wide bandwidth of this design is due to the application of GS in the SWS geometry, as well as the choice of operating point for beam-wave synchronism. As shown in Fig. \ref{fig_gain_vs_freq}, the smaller gain peak is at 68 GHz, corresponding to the point at which the slow space-charge wave of the beam is optimally velocity-synchronized to the guided forward even mode shown in Fig. \ref{fig_vph}.


\begin{figure}[!t]
\centering
\includegraphics[width=0.9\columnwidth]{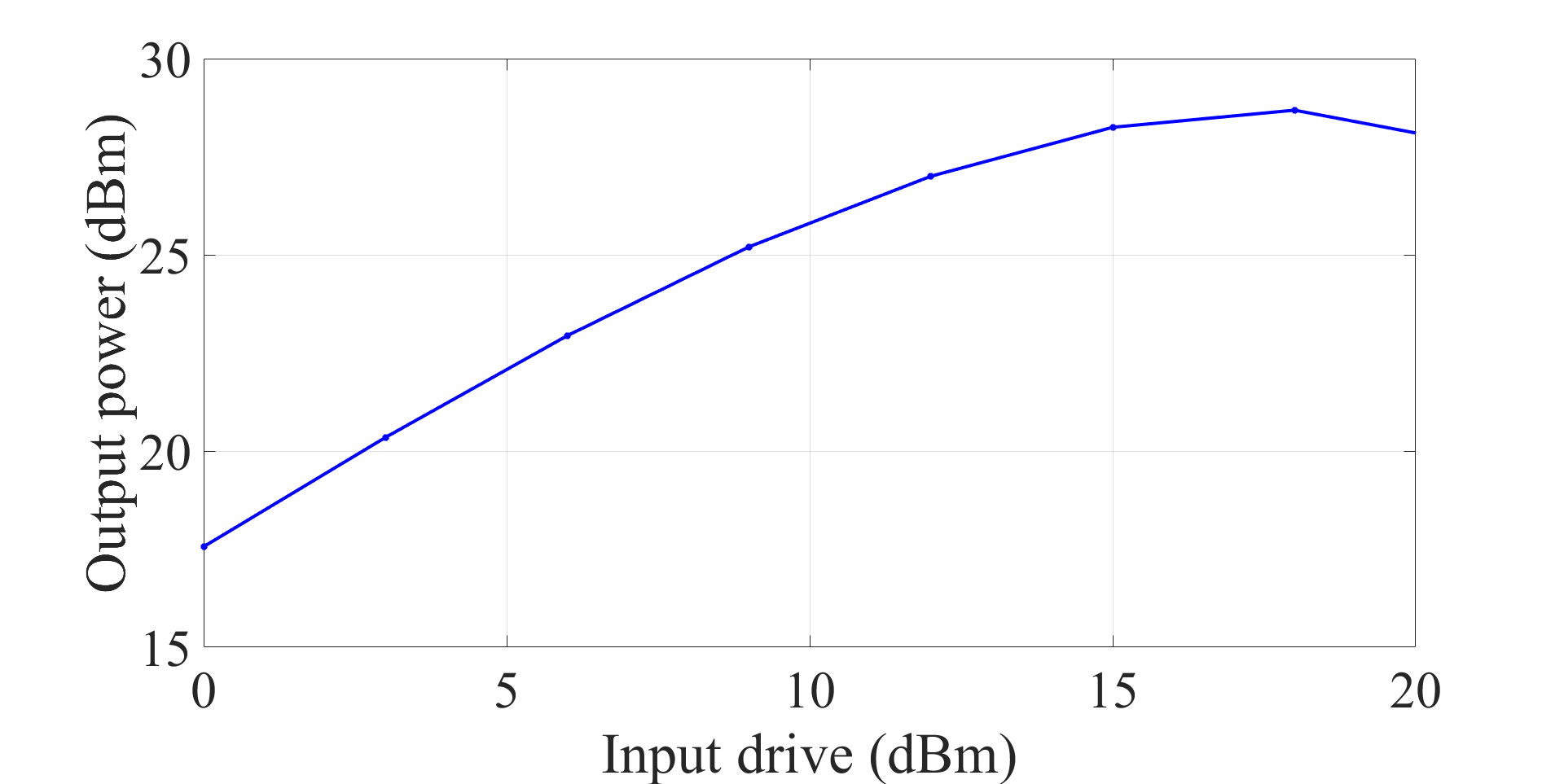}
\caption{Output power vs input drive for the TWT at a center frequency of 62 GHz.}
\label{fig_pout_vs_pin}
\end{figure}

Next, the output power vs input drive curves were obtained by sweeping the power of a 62 GHz CW excitation signal from 0 dBm to 20 dBm, as shown in Fig. \ref{fig_pout_vs_pin}. The saturation output power of the TWT is approximately 28.7 dBm (741 mW). The saturation power is small for several reasons: (i) the DC beam power is small (57.2 W) due to the small beam current used, (ii) the interaction impedance is smaller at the center frequency than it is close to the lower band-edge of the SWS, and (iii) the beam is optimally synchronized at a higher frequency than 62 GHz in order to achieve broadband operation. The basic DC-to-RF conversion efficiency of this TWT design is approximately 1.2\%.

To summarize, this work showcases a slow-wave structure suitable for sheet-beam TWTs which exhibits a 3-dB gain bandwidth of 55-68 GHz (22\% relative bandwidth centered at approximately 62 GHz) and a basic DC-to-RF conversion efficiency of 1.2\% using a 5.2 kV, 11 mA beam. For comparison, the GS double corrugated gap waveguide shown in \cite{saavedra2024wideband} exhibited a relative bandwidth of 23\% centered at 61 GHz and a DC-to-RF conversion efficiency of 3.1\% with a 21.8 kV, 25 mA pencil beam. Other kinds of wideband millimeter-wave GS sheet beam TWTs have also been demonstrated in the literature. In \cite{pershing2014demonstration}, a pulsed high-power sheet beam coupled-cavity TWT was demonstrated experimentally with a relative bandwidth of 14.7\% at a center frequency of 34 GHz and a DC-to-RF conversion efficiency of approximately 15\% using a high-current 20.5 kV, 3.5 A beam. For the more common staggered double vane sheet beam TWT design, a relative 3-dB bandwidth of approximately 25\% at a center frequency of 220 GHz and a DC-to-RF conversion efficiency of approximately 3\% to 5.5\% was achieved in \cite{shin2011modeling} using a 20 kV, 250 mA beam. 
While the relative bandwidth in \cite{shin2011modeling} appears to be comparable with the one in our work, it should be noted that input and output waveguide couplers were not included in \cite{shin2011modeling}. Including realistic waveguide couplers  will likely reduce the usable relative bandwidth of their device to something smaller than 25\%. For further comparison, in \cite{karetnikova2018gain}, a staggered vane TWT was shown with a relative 3-dB bandwidth of approximately 7.6\% centered at 185 GHz with a conversion efficiency of 3\% using a 20 kV, 100 mA beam. Compared to other wideband sheet beam TWTs in the literature, this work has a comparable or better relative bandwidth. However, the basic efficiency in this work is rather low due to the small beam current available from the proposed diamond FEA cathode. The efficiency of this design can be greatly improved by choosing a beam voltage for a more narrowband and a lower frequency operating point (where the interaction impedance is high) or by further increasing the beam current (increasing the beam area compression) to improve the Pierce gain parameter $C$, which, for small values, is proportional to the theoretical maximum efficiency of the TWT \cite{nordsieck1953theory}.

\section{Conclusion}
We have showcased a new wideband SWS geometry for millimeter-wave TWTs that makes use of GS to achieve a large relative 3 dB gain bandwidth of 22\%. The TWT shown in this work is designed to work with a sheet beam produced by a diamond FEA cathode and, while the basic efficiency and saturation power is poor due to the choice of operating point for beam-wave synchronization and the limited current density available from diamond FEA cathodes, the basic efficiency can be greatly improved by using a larger beam current or a more narrowband design.

\appendices
\section{Coupler Design and Cold Scattering Parameters}\label{sec_coupler}
\begin{figure}[!t]
\centering
\includegraphics[width=0.9\columnwidth]{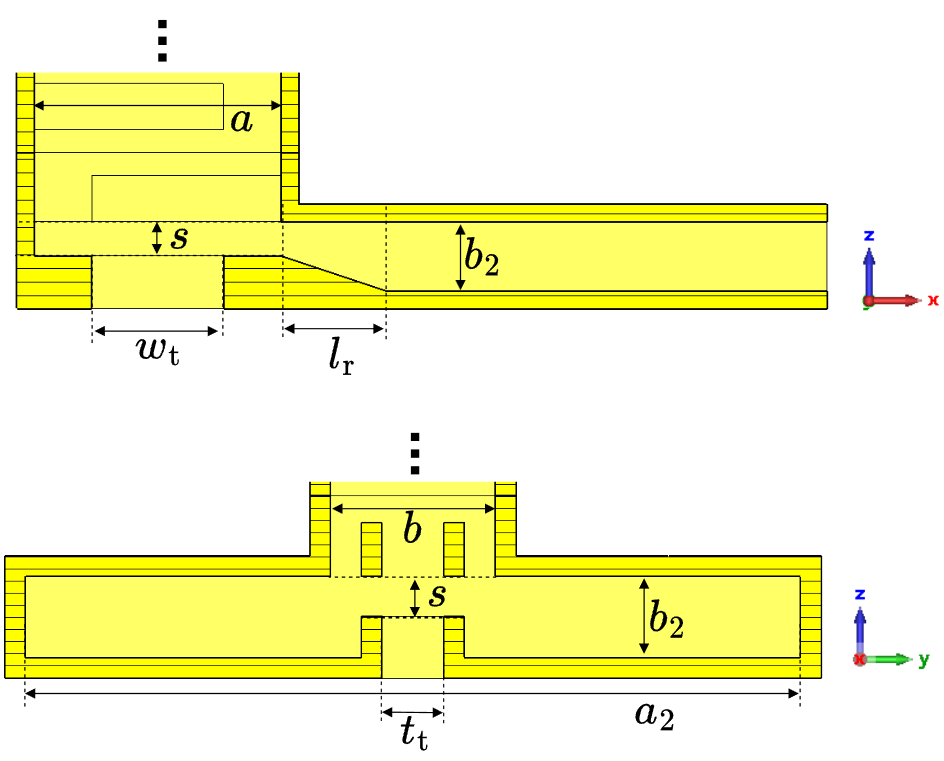}
\caption{Coupler design for the sheet-beam TWT input and output.}
\label{fig_coupler}
\end{figure}

The cross-sections of the coupler used for this design are shown in Fig. \ref{fig_coupler}. The coupler is designed to excite only the even mode in the SWS by symmetrically exciting the $E_z$ field on each pair of pillars. The TE10 mode of the larger rectangular waveguide used to excite the TWT and to extract power couples to the SWS through the SWS's $E_z$ field directly, without the need for any tapering of the pillars or cell period. The coupler also uses a ridged section to enhance the $E_z$ field coupling and match between the larger rectangular waveguide and the SWS. For simplicity, vacuum windows and height transitions for the rectangular waveguide sections of the coupler are not included in this study, but they will be necessary in practice.

Dimensions for the input and output waveguide couplers are provided in Table \ref{tab:coupler_dims}. The transverse width and height of the SWS ($a$ and $b$, respectively) are provided in Table \ref{tab:cell_dims}. The larger rectangular waveguide section has a width of $a_2$, corresponding to the WR-15 waveguide standard. The height of the rectangular waveguide section is $b_2$. Waveguide height tapering (to be able to connect the TWT ports to a standard WR15 waveguide) and the inclusion of vacuum windows will be necessary for laboratory testing, but such features are currently outside of the scope of this paper. The ridge section protrudes from a broad wall of the larger rectangular waveguide to create a gap spacing of $s$ between the ridge and the pair of pillars that are flush with the opposite broad wall. A ramp length of $l_{\rm{r}}$ is used to transition from the empty rectangular waveguide to the ridged waveguide. Inside the ridge, a beam tunnel hole is cut with dimensions of $w_{\rm{t}}$ and $t_{\rm{t}}$ to give adequate clearance for the sheet beam while also being in cutoff for the frequencies that we operate at.

\begin{figure}[!t]
\centering
\includegraphics[width=0.9\columnwidth]{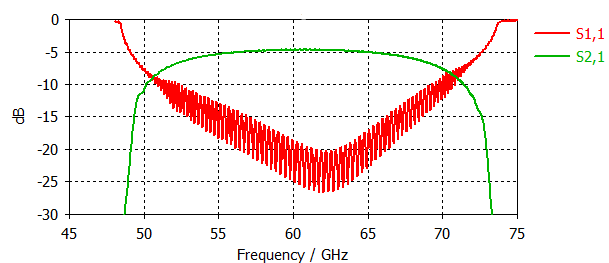}
\caption{Scattering parameters for the 60-cell cold SWS with copper walls.}
\label{fig_sparams}
\end{figure}

The cold scattering parameters of the finite-length 60 cell copper SWS,  illustrated in Fig. \ref{fig_finite_length}, are shown in Fig. \ref{fig_sparams}. There is good matching over the whole bandwidth of interest.

\begin{table}[]
\caption{Waveguide coupler dimensions}
\centering
\begin{tabular}{c||cccccc}
\hline
\textbf{Dim.} & \textbf{$a_2$} & \textbf{$b_2$} & \textbf{$w_{\rm{t}}$} & \textbf{$t_{\rm{t}}$} & \textbf{$s$} & \textbf{$l_{\rm{r}}$} \\ 
\hline
\textbf{Val. (mm)} & 3.76 & 0.39 & 0.75 & 0.3 & 0.19 & 0.6  \\ 
\hline
\end{tabular}
\label{tab:coupler_dims}
\end{table}

\section{Vanishing Backward Wave Interaction Impedance}\label{sec_BWO_impedance}
\begin{figure}[!t]
\centering
\subfloat(a){\includegraphics[width=0.9\columnwidth]{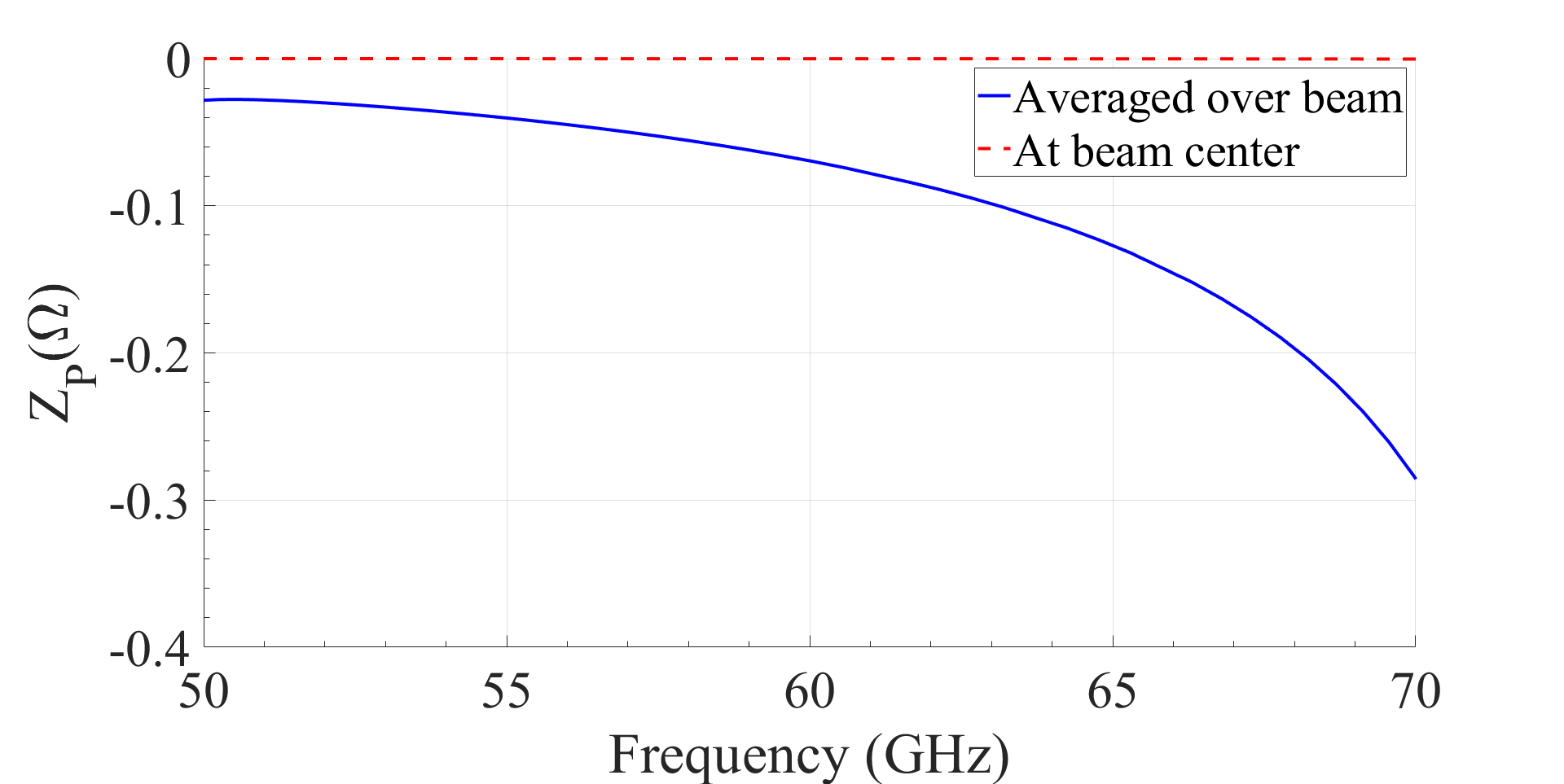}}\label{Fig:Zp_backward}
\subfloat(b){\includegraphics[width=0.9\columnwidth]{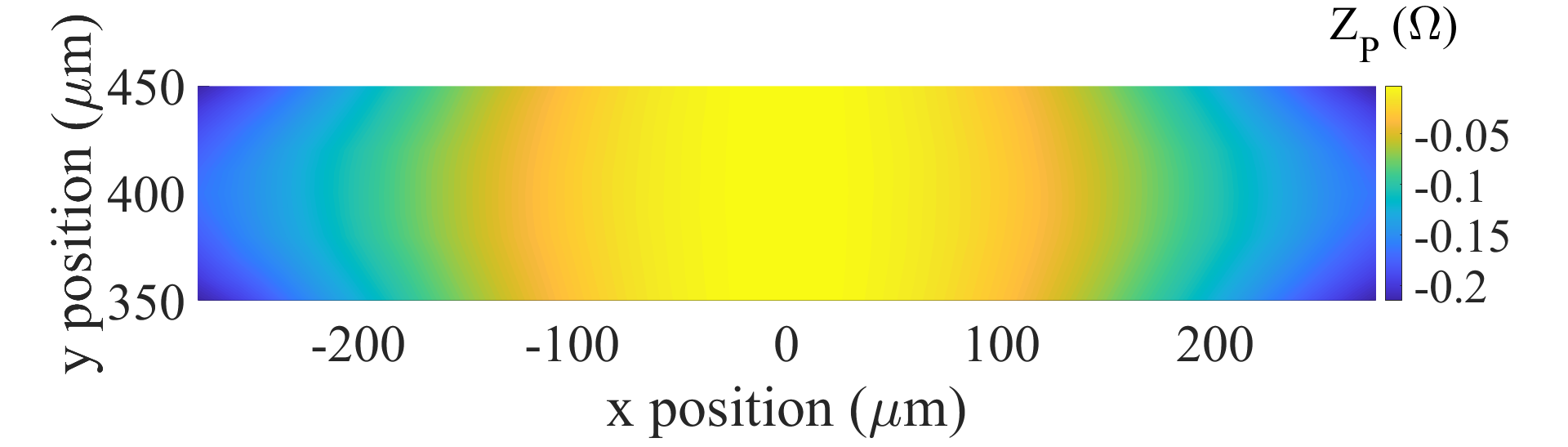}}\label{Fig:ZpContour_backward}
\caption{(a) The small interaction impedance for the {\em backward} even mode as a function of frequency averaged over the beam cross-section (solid blue) and at the center of the electron beam (dotted red). This latter vanishes because of GS topological properties. The averaged interaction impedance over the beam cross-section is very small compared to that of the forward wave in Fig. \ref{fig_Zp}. (b) Low local values of the interaction impedance of the backward even mode over the beam cross-section at 61 GHz, i.e., near the $\beta_nd=3\pi$ point. }
\label{fig_Zp_backward}
\end{figure}

As was also demonstrated in \cite{castro2023wide} for a similar SWS topology, the on-axis electric field corresponding to the {\em backward} even mode between the first and second space harmonics (i.e., in the interval $\beta_n d/\pi \in [2,4]$) has a negligible $E_z$-component in its wavenumber spectrum. The {\em forward} even mode instead has a significant $E_z$ polarization. This topological property is caused by GS and it is very useful in the design of a TWT. As a consequence, the on-axis interaction impedance for the backward wave, as calculated in Eqn. (\ref{eqn:interaction_impedance}) is effectively $0~\Omega$, as shown in Fig. \ref{fig_Zp_backward}. Near the edges of the beam, the interaction impedance for the backward even mode is still present, but it is small relative to the interaction impedance of the forward even mode shown in Fig. \ref{fig_Zp}. The negative sign for $Z_{\mathrm{P}}$ in Fig. \ref{fig_Zp_backward} is due to the negative sign of the power flow in Eqn. (\ref{eqn:interaction_impedance}), simply indicating that power flows in the $-z$ direction for the backward wave. This $E_z \approx 0$ property of the backward wave, induced by GS, makes the TWT less susceptible to backward-wave oscillations.

\section{Considerations for Assembly}\label{sec_assembly}
\begin{figure}[!t]
\centering
\includegraphics[width=0.9\columnwidth]{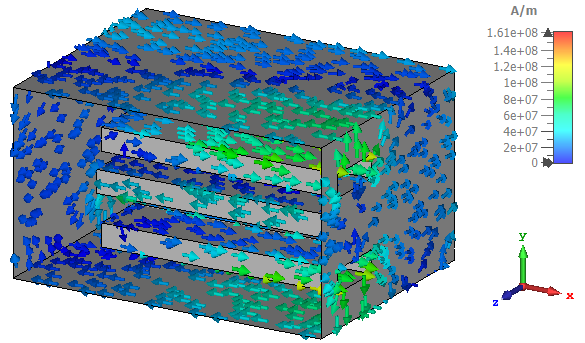}
\caption{Surface currents of the forward even mode in the unit cell of the SWS at 60 GHz, slightly below $\beta_nd=3\pi$. Near the center of the narrow lateral walls, surface currents only run along the $z$-direction}
\label{fig_surf_current}
\end{figure}

The surface currents on the walls of the SWS for the forward mode slightly below the $\beta_nd=3\pi$ operating point are shown in Fig. \ref{fig_surf_current}. On the broad top and bottom walls of the surrounding waveguide, there are significant surface currents  along the $x$ direction. Whereas, on the narrow lateral walls of the surrounding waveguide, the surface currents near the center of the walls mostly run along the $z$ direction. If the SWS is assembled by welding two halves together along the $y-z$ plane, there may be significant losses where the surface currents cross the weld seam. Thus, it is preferable to assemble the structure with weld seams along the $x-z$ plane. The structure can be assembled using layered metal plates, like was done in \cite{wang2024investigation,wang2024staggered}.

\section*{Acknowledgment}
This material is based upon work supported by the UCI-LANL graduate fellowship program, and the Air Force Office of Scientific Research MURI award number FA9550-20-1-0409 administered through the University of New Mexico. This work was also supported by the U.S. Department of Energy,
through the Laboratory Directed Research and Development
(LDRD) Program at Los Alamos National Laboratory, under
project number 20230110ER, LA-UR-25-20407. The authors are thankful to DS Simulia for providing CST Studio Suite that was instrumental in this study.

\ifCLASSOPTIONcaptionsoff
  \newpage
\fi



\bibliographystyle{IEEEtran}

\end{document}